\documentclass[preprint,showkeys]{revtex4}
\usepackage{graphicx}
\usepackage{epsfig}
\usepackage{amssymb}
\usepackage{color}
\usepackage{amsmath}
\usepackage{lipsum}
\usepackage[font={small}]{caption}
\usepackage{graphics}
\usepackage{tensor}
\usepackage{filecontents}
\usepackage{epsf}
\usepackage{psfrag}
\usepackage{epstopdf}
\usepackage{graphicx}
\usepackage{caption}
\usepackage{epstopdf}
\usepackage{epsfig}

\begin{document}

\title{Exploring Modifications to FLRW Cosmology with General Entropy and Thermodynamics: A new Approach}
\author{ A. Khodam-Mohammadi$^{1}$\footnote{Email: Khodam@basu.ac.ir} and M. Monshizadeh$^{2}$}
\affiliation{$^{1}$Department
of Physics, Faculty of Science, Bu-Ali Sina University, Hamedan
65178, Iran\\$^{2}$Physics Department, Faculty of Science,
Islamic Azad University of Hamedan, Hamedan 65181, Iran}

\begin{abstract}
The investigation of modifications to the FLRW cosmology resulting from the consideration of a general entropy for the cosmological apparent horizon is the subject of this study. Building upon the work of Nojiri and collaborators in 2022, who introduced a class of generalized entropies with four parameters capable of converging to familiar entropies and addressing specific cosmological issues, our research explores the impact of correcting the entropy on the energy-momentum tensor of the cosmic fluid from the outset. Our calculations demonstrate that, by employing a correction function $f(\rho)$ to modify the energy-momentum density tensor, the entropic area law (Bekenstein-Hawking entropy) can still be regarded as a general entropy. The construction of the function $f(\rho)$ is facilitated through considerations of the thermodynamics associated with the apparent horizon. Additionally, we investigate the first and second laws of thermodynamics within this framework and illustrate how the limitations imposed on the equation of state of the cosmic fluid can be resolved through the incorporation of this correction function. Finally, we compute cosmography parameters to analyze the kinematics of the universe, with particular attention given to the notable influence of the correction function $f(\rho)$ on these parameters. This paper provides valuable insights into the application of general entropies to the apparent horizon of the universe.
\end{abstract}


\maketitle 

\section{Introduction}
Gravity and thermodynamics are two fundamental areas of physics that have traditionally been studied independently. However, the exploration of black hole mechanics, particularly the discovery of Bekenstein-Hawking entropy ($S_{BH}$) \cite{Bekenstein:1973ur, Hawking:1975vcx}, has sparked a growing interest in understanding the connection between these two fields \cite{Jacobson:1995ab, Hayward:1997jp, Padmanabhan:2003gd, Padmanabhan:2009vy}. It has been established that Einstein's equation can be interpreted as the first law of thermodynamics \cite{Padmanabhan:2013nxa}, and the development of FLRW cosmology based on Einstein's field equation further solidifies the analogy between thermodynamics and cosmology. 

The apparent horizon, a specific type of cosmological horizon, plays a crucial role in this study. It represents a boundary beyond which light cannot escape to infinity and is causally connected to an observer in spacetime. Remarkably, the apparent horizon can be treated as a thermodynamic system with temperature ($T$) and entropy ($S$). This concept was first proposed by Jacobson in 1995 \cite{Jacobson:1995ab}, who demonstrated that the Einstein field equation could be derived from the laws of thermodynamics. The temperature is determined by the surface gravity of the horizon, while the entropy is proportional to the area of the horizon in Planck units. By studying the thermodynamics of the FLRW apparent horizon, researchers can gain valuable insights into various features and challenges in this field of research \cite{Sanchez:2022xfh}.

In addition to black hole entropy, alternative entropy functions have been proposed, such as Tsallis entropy \cite{Tsallis:1987eu}, R\'{e}nyi entropy \cite{Renyi:1960}, Barrow entropy \cite{Barrow:2020tzx}, Sharma-Mittal entropy \cite{SayahianJahromi:2018irq}, Kaniadakis entropy \cite{Kaniadakis:2005zk, Drepanou:2021jiv}, and loop quantum gravity entropy \cite{Majhi:2017zao, Liu:2021dvj}. These entropies are all monotonically increasing functions of $S_{BH}$, satisfy the third law of thermodynamics ($S\rightarrow 0$ as $S_{BH}\rightarrow 0$), and converge to $S_{BH}$ for suitable parameter choices and some considerations \cite{Nojiri:2021czz}.

Recently, generalized entropies with three, four, and six parameters have been proposed, some of which can reproduce known entropies by selecting appropriate parameters. These generalized entropies have shown promise in addressing problems in various cosmological areas, including black holes and bounce cosmology \cite{Nojiri:2022aof, Nojiri:2022dkr, Odintsov:2022qnn, Odintsov:2023qfj,Nojiri:2022sfd}. Notably, in \cite{Sheykhi:2018dpn, Nojiri:2022nmu}, the authors generalized the entropy beyond $S_{BH}$ and found that it necessitates modifications to the formal Friedmann equations. They also made an intriguing conjecture stating that the minimum number of parameters required in a generalized entropy function capable of encompassing all the aforementioned known entropies is four \cite{Odintsov:2023qfj}.

The modification of the Friedmann equations has further fueled our curiosity to delve deeper into this subject. Does the generalization of entropy lead to changes in the energy-momentum density tensor of the cosmic flux? If so, does it also necessitate modifications to the continuity equation? And

if these changes occur, will the laws of thermodynamics still hold?

Furthermore, in the study of the universe's kinematics, cosmologists examine various cosmographic parameters, such as the Hubble parameter, deceleration parameter, jerk parameter, snap parameter, and more \cite{Visser:2004bf, Ivashtenko:2019aqf, Rezaei:2020lfy, Mehrabi:2021cob}. How do generalized entropies affect these cosmographic parameters?

This essay aims to address these questions. In Section \ref{Sec2}, we will provide a brief overview of the method presented by Nojiri et al. \cite{Nojiri:2022nmu} regarding the modification of the Friedmann equations resulting from the application of generalized entropy. Then, in Section \ref{Sec3}, we propose that the same changes achieved in the previous section can be obtained by modifying the energy-momentum density of the perfect cosmic fluid using a correction function. However, in this case, the continuity equation needs to be modified as well. Section \ref{Sec4} explores the reconstruction of the correction function applied to the energy-momentum tensor using the first law of thermodynamics and a four-parameter generalized entropy. The second law of thermodynamics is discussed in Section \ref{Sec5}, and the impact of applying generalized entropies on cosmographic parameters is investigated in Section \ref{Sec6}. Finally, the work concludes with a summary and some concluding remarks.

Throughout this article, I use natural units where `$c=G=\hbar=1$'. The over-dot notation denotes a derivative with respect to time ($d/d t$), while the prime notation represents a derivative with respect to the function's arguments.
     
\section{Modification of FLRW cosmology by considering a general entropy}\label{Sec2}
In this section, we provide a brief review of the method presented in \cite{Nojiri:2022nmu}.

In an isotropic FLRW metric, the line element is given by
\begin{equation}
ds^2=h_{ab}dx^a dx^b+R^2(d\theta^2+\sin^2\theta d\phi^2),
\end{equation}
where $x^0=t$, $x^1=r$, $R=a(t)r$, and $h_{ab}=\text{diag}\left(-1,~\frac{a(t)^2}{1-kr^2}\right)$ with its determinant denoted as `$h$'. The Bekenstein-Hawking entropy, radius, surface gravity, and temperature of the apparent horizon are given by \cite{Sanchez:2022xfh,Cai:2005ra,Binetruy:2014ela,Hayward:1997jp}:
\begin{eqnarray}
S_{BH}&=&\frac{A}{4}=\pi R_h^2, \\
R_h^2&=&\frac{1}{H^2+\frac{k}{a^2}}, \\
\kappa&=&\frac{1}{2\sqrt{-h}}\frac{\partial}{\partial x^a}(\sqrt{-h}h^{ab}\frac{\partial R_h}{\partial x^b}), \\
T_h&=&\frac{|\kappa|}{2\pi}=\frac{1}{2\pi R_h}\left|1-\frac{\dot{R_h}}{2 H R_h}\right|.
\end{eqnarray}

Following the recent paper by Nojiri et al. \cite{Nojiri:2022nmu}, a four-parameter general entropy was proposed as
\begin{equation}
S_G(\alpha_+,\alpha_-,\gamma,\beta)=\frac{1}{\gamma}\left[\left(1+\frac{\alpha_+ S_{BH}}{\beta}\right)^\beta-\left(1+\frac{\alpha_- S_{BH}}{\beta}\right)^{-\beta}\right]. \label{SG}
\end{equation}

Using the first law of thermodynamics, $dE=-du=-\delta Q-p dV$, where $\delta Q=T dS$ represents heat transfer from the apparent horizon and $p$ is the pressure of the fluid in the volume $V$, the following quantities were calculated:
\begin{eqnarray}
&&E=\frac{4}{3}\pi R^3_h \rho,~~~~~ V=\frac{4}{3}\pi R^3_h, \\
&&\frac{dE}{dt}+p\frac{dV}{dt}=-T\frac{dS}{dt}=\frac{4\pi}{3}\dot{\rho}R^3_h\left[1+3\frac{\dot{R_h}}{R_h}\frac{p+\rho}{\dot{\rho}}\right], \notag \\
&&\frac{-1}{2\pi R_h}\left|1+\frac{R^2_h}{2}(\dot{H}-\frac{k}{a^2})\right|\dot{S}=\frac{4\pi}{3}\dot{\rho}\left[1-R^2_h(\dot{H}-\frac{k}{a^2})\right]. \label{sdot}
\end{eqnarray}

In these equations, the formal continuity equation $\dot{\rho}=-3H(\rho+p)$ is used. By considering the general entropy as a function of $S_G(S_{BH})$, we obtain   
\begin{equation}
\dot{S}=-2\pi R^4_h H (\dot{H}-\frac{k}{a^2})S'({S_{BH}}).\label{sprim}
\end{equation}

Inserting (\ref{sprim}) into (\ref{sdot}) leads to
\begin{equation}
\left[1+\frac{\frac{3R_h^2}{2}(\dot{H}-\frac{k}{a^2})}{1-R_h^2(\dot{H}-\frac{k}{a^2})}\right] S'(\dot{H}-\frac{k}{a^2})=-4\pi(\rho+p), \label{sprime2}
\end{equation}
which has the capability to produce a modified version of the Friedmann equation. Although solving the differential equation (\ref{sprime2}) is challenging when considering the four parameters of general entropy (\ref{SG}), by considering a Tsallis limit of the general entropy ($\alpha_-=0$ and $\alpha_+\rightarrow \infty$), the following expression was obtained \cite{Nojiri:2022nmu}:
\begin{equation}
S_T=S_0\left(\frac{S_{BH}}{S_0}\right)^\delta,
\end{equation}
where $S_0$ is a constant and $\delta$ is the exponent of the Tsallis entropy. The differential equation (\ref{sprime2}) was solved under this assumption, resulting in
\begin{equation}
H^2+\frac{k}{a^2}\sim \pi \left\{\frac{8}{3}\frac{2-\delta}{\delta}S_0^{\delta-1}(\rho+\Lambda)\right\}^{\frac{1}{2-\delta}},
\end{equation}
where $\Lambda$ is a constant of integration that may be interpreted as the cosmological constant. One can rewrite this equation as 
\begin{equation}
H^2+\frac{k}{a^2}\sim \frac{8\pi}{3}\rho f(\rho), \label{fried1}
\end{equation}
where 
\begin{equation}
f(\rho)=\frac{3}{8\rho} \left\{\frac{8\pi}{3}\frac{2-\delta}{\delta}S_0^{\delta-1}(\rho+\Lambda)\right\}^{\frac{1}{2-\delta}}. \label{fp}
\end{equation}
Considering relation (\ref{fried1}), we propose the following conjecture:

Conjecture: ``\textit{By generalizing entropy in an FLRW cosmology, the energy-momentum density tensor of the cosmic fluid is modified in such a way that a common correction function $f(\rho)$ multiplies both the energy density and pressure terms. This modification guarantees the preservation of the perfectness property and the equation of state parameter of the cosmic fluid.}"

In the following, we examine this conjecture and attempt to reconstruct $f(\rho)$ by considering the general entropy.

\section{modification of energy-momentum density tensor corresponding to general entropy}\label{Sec3}
In the previous section, a conjecture was proposed, suggesting that the general entropy could potentially modify the energy-momentum density tensor through a multiplication correction function.

By considering the following modified energy-momentum tensor for a perfect fluid:
\begin{equation}
T^\mu_\nu=diag(-\rho f(\rho), p f(\rho), p f(\rho), p f(\rho)),
\end{equation}
where the equation of state parameter (EoS) is defined as $w=p/\rho$, the conservation law $T^{\mu\nu}_{;\nu} =0$ leads to the following equation:
\begin{eqnarray}
T^{\mu\nu}_{;\nu} =0 &\Rightarrow a^3\frac{d (p
	f(\rho))}{dt}=\frac{d}{dt}[a^3(\rho+p)f(\rho)]\Rightarrow
a^3[\dot{p} f(\rho)+f^\prime(\rho)\dot{\rho}p]= \notag\\
&3a^2\dot{a}[\rho+p]f(\rho)+a^3(\dot{\rho}+\dot{p})f(\rho)+a^3(\rho+p)f^\prime(\rho)\dot{\rho}.\notag
\end{eqnarray}
After a little simplification, we find
\begin{equation}
\dot{\rho}[1+\rho\frac{f^\prime(\rho)}{f(\rho)}]+3H(\rho+p)=0.\label{consf}
\end{equation}
For instance, when considering the Tsallis entropy and Eq. (\ref{fp}), the modified continuity equation becomes:
\begin{equation}
\dot{\rho}+3H(\rho+p)(2-\delta)=0\label{cont-eq},
\end{equation}
which differs from the continuity equation discussed in the previous section.

By taking the time derivative of Eq. (\ref{fried1}) and utilizing Eq. (\ref{consf}), the second modified Friedmann equation can be derived as follows:
\begin{equation}
\dot{H}-\frac{k}{a^2}=-4 \pi (\rho+p) f(\rho).\label{FR2}
\end{equation}

\section{Reconstruction $f(\rho)$ by considering the first law of thermodynamics}\label{Sec4}
Using the modified Friedmann equations (\ref{fried1}) and (\ref{FR2}), along with the first law of thermodynamics, $dE=-\delta Q+W d V$, where $W=-1/2 T^{a b}h_{a b}$ represents the work density and $\delta Q=T d S$, we can derive the following relations. Additionally, we utilize the equations:
\begin{equation}
E=\frac{4}{3}\pi R^3_h \rho f(\rho),~~ V=\frac{4}{3}\pi R^3_h,~~ W=-\frac{1}{2}(p-\rho) f(\rho),
\end{equation}
By doing so, we obtain the following relations:
\begin{eqnarray}
&&\frac{d E}{d t}-W \frac{d V}{d t}=\frac{4 \pi}{3}\dot{\rho}R^3_h f(\rho)[1+\frac{\rho f^\prime (\rho)}{f(\rho)}+\frac{3}{2}\frac{\dot{R_h}}{R_h}\frac{p+\rho}{\dot{\rho}}]\notag\\
&&-T \frac{d S}{d t}=\frac{4 \pi}{3}\dot{\rho}R^3_h f(\rho)[1+\frac{R^2_h}{2}(\dot{H}-\frac{k}{a^2})][1+\frac{\rho f^\prime (\rho)}{f(\rho)}]\Rightarrow \notag \\
&&\frac{-1}{2\pi R_h}|1+\frac{R^2_h}{2}(\dot{H}-\frac{k}{a^2})|\dot{S}=\frac{4 \pi}{3}\dot{\rho}R^3_h f(\rho)[1+\frac{R^2_h}{2}(\dot{H}-\frac{k}{a^2})] [1+\frac{\rho f^\prime (\rho)}{f(\rho)}]. \label{aa}
\end{eqnarray}
Furthermore, by using
\begin{equation}
\dot{S}=8\pi^2 R^4_h H (\rho+p)f(\rho) S^\prime(\tilde{S}_{BH}),~\&~\dot{\rho}=\frac{-3 H (\rho+p)}{1+\frac{\rho f^\prime (\rho)}{f(\rho)}},
\end{equation}
where $\tilde{S}_{BH}=\frac{3}{8\rho f(\rho)}$ is the Bekenstein-Hawking entropy in this new point of view, the relation (\ref{aa}) leads to
\begin{equation}
S^\prime(\tilde{S}_{BH})=1.
\end{equation}
In the case of standard FLRW cosmology with $f(\rho)=1$, this yields $S_G=S_{BH}+C$, which is as expected if the integration constant $C=0$. In order to reconstruct the correction function $f(\rho)$ from this perspective, we need to compare the correspondence between these two approaches. Let's consider:
\begin{eqnarray}
\tilde{S}_{BH}=S_G(S_{BH}),
\end{eqnarray}
where $S_G$ represents the generalized entropy (\ref{SG}). As a result, we obtain:
\begin{equation}
\frac{3}{8\rho f(\rho)}=\frac{1}{\gamma}[(1+\frac{3\alpha_+ }{8\beta\rho})^\beta-(1+\frac{3\alpha_- }{8\beta\rho})^{-\beta}].
\end{equation}
Simplifying further, we find that:
\begin{equation}
f(\rho)=\frac{3\gamma}{8\rho}[(1+\frac{3\alpha_+ }{8\beta\rho})^\beta-(1+\frac{3\alpha_- }{8\beta\rho})^{-\beta}]^{-1}.
\end{equation}
With this approach, we can easily determine the correction function corresponding to any appropriate entropy function in the work and incorporate it into the tensor of the energy-momentum density of the cosmic fluid. The subsequent steps of the process remain the same as standard cosmological calculations. It is important to note that in this method, the perfect fluid condition is still valid and the equation of state (EoS) of the cosmic fluid remains unchanged.

Table \ref{chart} provides several correcting functions $f(\rho)$ corresponding to different entropy functions $S_G$.

\begin{table*}[h]
	\begin{center}
		\begin{tabular}{|r|c|c|c|l|}
			\hline
			
			&Entropy & $S_G$ & $f(\rho)$ \\
			\hline
			1 &Tsallis & $S_0(\dfrac{S_{BH}}{S_0})^\delta$ & $(\dfrac{8}{3}S_0\rho)^{\delta-1}$ \\
			\hline
			2 &R\'{e}nyi & $ \dfrac{1}{\alpha}\ln(1+\alpha S_{BH})$ & $\dfrac{3\alpha}{8\rho}[\ln(1+\dfrac{3\alpha}{8\rho})]^{-1} $ \\
			\hline
			3 &Sharma-Mittal &$\dfrac{1}{R}[(1+\delta S_{BH})^{R/\delta}-1]$ & $\dfrac{3R}{8\rho}[(1+\dfrac{3\delta}{8\rho})^{R/\delta}-1]^{-1} $ \\
			
			\hline
			
			4 &Kaniadakis &$ \dfrac{1}{K}\sinh(KS_{BH})$ & $\dfrac{3K}{8\rho}[\sinh(\dfrac{3K}{8\rho})]^{-1} $\\
			
			\hline
			
			5 &Loop quantum gravity&$ \dfrac{1}{1-q}\{\exp[(1-q)\Lambda(\gamma_0)S_{BH}]-1\}$ &$\dfrac{3(1-q)}{8\rho}\{\exp[\dfrac{3}{8\rho}(1-q)\Lambda(\gamma_0)]-1\}^{-1} $\\
			\hline
			
		\end{tabular}
	\end{center}
	\caption{Reconstruction of $f(\rho)$ for various entropy functions}. \label{chart}
\end{table*}

Another outcome of this analysis is that by comparing the two forms of the first law of thermodynamics in sections II and IV, we deduce that:
\begin{equation}
-W=p\Rightarrow \frac{1}{2}(p-\rho)f(\rho)=p\Rightarrow~w=(1-\frac{2}{f(\rho)})^{-1}.\label{EoS1}
\end{equation}
Hence, when considering the general entropy and the consequent modification of the stress tensor, we find that although the standard FLRW cosmology assumes $S_G=S_{BH}$ and $f(\rho)=1$, implying that only the equation of state (EoS) of the cosmic fluid should be considered under the first law of thermodynamics with $w=-1$, the generalized entropy prescription may allow for a wider range of matter with a dynamic EoS. Thus, the first law of thermodynamics permits a broader range of matter in the generalized entropy prescription compared to standard FLRW cosmology. It should be noted that equation (\ref{EoS1}) indicates a singularity in the EoS parameter at a certain point, but this singularity's location can be altered by selecting appropriate parameters. However, it is important to emphasize that this equation of state function was derived solely to satisfy the first law of thermodynamics. Therefore, it only indicates the range in which the first law is satisfied and does not impose any particular value.

An important point to highlight in this regard is the power and simplicity of the new method for modifying the Friedmann equations compared to the previous method discussed in Section II. Additionally, in the future, by considering other general entropies such as non-singular entropy or entropy with additional parameters, this modification process can be easily carried out. Moreover, in this method, all kinds of entropies can be accommodated by the modified Bekenstein-Hawking entropy "$\tilde{S}_{BH}=3/(8\rho f(\rho))$" with its corresponding $f(\rho)$.

\section{second law of thermodynamics}\label{Sec5}
In the second law of thermodynamics, the entropy must be a convex extensive function of the system's state variables. Therefore, we have the conditions $\dot{S}\geqslant 0$ and $\ddot{S}\leqslant 0$ \cite{Sanchez:2022xfh}. From the equation $S_G=\tilde{S}_{BH}$ and the modified continuity equation, we obtain
\begin{eqnarray}
\dot{S}&=&3 H S_G (1+w)\geqslant 0 \label{Sd}\\
\ddot{S}&=&\dot{S}[3H(1+w)+\frac{\dot{H}}{H}+\frac{\dot{w}}{1+w}]\leqslant 0. \label{Sdd}
\end{eqnarray}

The first relation, (\ref{Sd}), implies that $w\geqslant-1$. Moreover, Equation (\ref{Sdd}) gives:

\begin{equation}
3H(1+w)+\frac{\dot{H}}{H}+\frac{\dot{w}}{1+w}\leqslant 0 \label{Sdd2}
\end{equation}

In Equation (\ref{Sdd2}), the first term is positive, while the remaining terms may be negative. In a non-phantom universe, where $\dot{H}<0$, the condition (\ref{Sdd2}) can still hold even if $\dot{w}$ becomes negative. In this case, we have:

\begin{equation}
\dot{H}\leqslant -3H^2(1+w)-\frac{H\dot{w}}{1+w}. \label{non-phantom-condition}
\end{equation}

In summary, by using the general entropy, it is possible to satisfy both the first and second laws of thermodynamics over a broad range of equations of state (EoS) parameters, provided that an appropriate set of parameters is chosen.

\section{The Effect of $f(\rho)$ on Cosmography Parameters}\label{Sec6}

Cosmography, or kinematics parameters, have proven to be useful tools for studying the kinematical state of our universe. These parameters are defined as follows \cite{Ivashtenko:2019aqf,Visser:2004bf}:

\begin{eqnarray}
H&=&\frac{\dot{a}}{a}\notag\\
q&=&-\frac{1}{H^2}\frac{\ddot{a}}{a}=-1-\frac{\dot{H}}{H^2}\notag\\
j&=&\frac{1}{H^3}\frac{a^{(3)}}{a}=1+3\frac{\dot{H}}{H^2}+\frac{\ddot{H}}{H^3}\notag\\
s&=&\frac{1}{H^4}\frac{a^{(4)}}{a}=1+6\frac{\dot{H}}{H^2}+4\frac{\ddot{H}}{H^3}+\frac{H^{(3)}}{H^4}+3\frac{\dot{H}^2}{H^4}\notag\\
&& .\notag\\
&& .\notag\\
&=&\frac{1}{H^n}\frac{a^{(n)}}{a}\label{cosmopar}
\end{eqnarray}

Here, $H$, $q$, $j$, and $s$ represent the Hubble, deceleration, jerk, and snap parameters, respectively. The superscript $(n)$ denotes the n'th derivative of a function with respect to cosmic time. In some works, higher time derivatives of parameters may also be needed \cite{Rezaei:2020lfy}. Using the notation $H^{(n)}$, the above relations can be rewritten as:

\begin{eqnarray}  
\dot{H}&=&-H^2(1+q)\notag\\
\ddot{H}&=&H^3(j+3q+2)\notag\\
H^{(3)}&=&H^4[s-4j-3q(q+4)-6]
\end{eqnarray}

Now, by considering the modified Friedmann equations (\ref{fried1}), (\ref{consf}), and (\ref{FR2}), we can obtain the following expressions:

\begin{eqnarray}
H^2&=&\frac{8\pi}{3}\rho f(\rho)-\frac{k}{a^2}\notag\\
\dot{H}&=&-4\pi\rho (1+w)f(\rho)+\frac{k}{a^2}\notag\\
\ddot{H}&=&12\pi\rho (1+w)^2 H f(\rho)-4\pi\rho \dot{w}f(\rho)-\frac{2 H k}{a^2}\notag\\
H^{(3)}&=&36\pi H (1+w)\rho f(\rho)[\dot{w}-H(1+w)^2]-48\pi^2 \rho^2 (1+w)^3 f(\rho)^2\notag\\
&&-4\pi\rho f(\rho)\ddot{w}+\frac{4k}{a^2}[H^2+3\pi\rho(1+w)^2f(\rho)+2\pi\rho(1+w)f(\rho)]-\frac{2k^2}{a^4}
\end{eqnarray}

These relations show that the correction function $f(\rho)$ affects all the cosmography parameters.

\section{Conclusion}
This study investigated the possible modifications to FLRW cosmology by incorporating a 4-parameter general entropy model for the universe's apparent horizon. Based on the work of Nojiri and colleagues in 2022, we proposed that the general entropies alter the cosmic fluid and the energy-momentum density tensor from the outset, leading to modifications of the Friedman equations that can be corrected with a function $f(\rho)$. We also proved that the continuity equation is modified, and the general entropy still behaves like a Bekenstein-Hawking entropy, $\tilde{S}_{BH}$ (entropic area law). The correction function $f(\rho)$ can be determined by equating the general entropy with the $\tilde{S}_{BH}$. A comprehensive table enumerating the corresponding corrected functions $f(\rho)$ associated with various known entropies, including Tsallis, Rényi, Sharma-Mital, Kaniadakis, and loop quantum gravity was prepared. Furthermore, we showed that, unlike standard cosmology, which restricts the cosmic fluid to $w=-1$, the cosmic fluid can contain a wider range of cosmic materials, satisfying both the first and second laws of thermodynamics. We also examined the kinematics of the universe by calculating the cosmography parameters and highlight the significant role of the correction function $f(\rho)$ in this analysis.

In conclusion, this paper provides important insights into the use of general entropies in cosmology and emphasizes the need to consider alternative entropy models for a deeper understanding of the universe. This new perspective could have significant implications for future research on the thermodynamics and evolution of the universe.
\bibliography{refth} 
\end{document}